\documentclass[a4paper,12pt,preprint,tightenlines,floatfix,showpacs,preprintnumbers,
nofootinbib,eqsecnum]{article}

\usepackage[utf8]{inputenc}
\usepackage{amssymb,amsmath}
\usepackage[english]{babel}

\usepackage[a4paper, total={7in, 9in}]{geometry}
\usepackage{wrapfig}
\usepackage{epstopdf}
\usepackage{graphicx,wrapfig}
\usepackage{feynmp}
\usepackage{feynman}
\usepackage{multirow}
\usepackage{mathtools}
\usepackage{tabularx}
\usepackage{makecell}
\usepackage{caption}
\usepackage[colorlinks,citecolor=blue,urlcolor=blue,bookmarks=false
]{hyperref}
\RequirePackage{flushend}
\RequirePackage[numbers,sort&compress]{natbib}
\usepackage{verbatim}
\usepackage{authblk}
\usepackage[font=footnotesize,labelfont=bf,singlelinecheck=false]{caption}
\usepackage[binary-units]{siunitx}
\usepackage{tikz}
\usetikzlibrary{arrows,shadows}
\usepackage{setspace}

\newcommand{\msbar}{\overline{\rm{MS}}}

\pdfoutput=1

\begin{document}
\pagenumbering{gobble}
\date{\today}

\title{\vspace{-1.5cm}\begin{flushright}
\small{INR-TH-2022-003}
\end{flushright} \vspace{1.5cm} \textbf{Notes on interplay of the QCD and   EW   perturbative corrections 
to the pole-running top-quark mass  ratio
}}

\author[1]{ A.~L.~Kataev\footnote{kataev@ms2.inr.ac.ru}}
\author[1, 2, 3, 4]{ V.~S.~Molokoedov\footnote{viktor\_molokoedov@mail.ru}}

\affil[1]{{\footnotesize Institute for Nuclear Research of the Russian Academy of Science,
 117312, Moscow, Russia}}
\affil[2]{{\footnotesize Research Computing Center, Moscow State University, 119991, Moscow, Russia}}
\affil[3]{{\footnotesize  Moscow Center for Fundamental and Applied Mathematics, 119992, Moscow, Russia}}
\affil[4]{{\footnotesize Moscow Institute of Physics and Technology, 141700, Dolgoprudny, Moscow Region, Russia}}

\maketitle

\begin{abstract}

A specific representation of the known one-loop EW correction to the relation between the pole and running $\msbar$-scheme masses of the top-quark through particle masses of the Standard Model is given within the Fleischer-Jegerlehner tadpole scheme, where the vacuum expectation value of the Higgs field is renormalized. The importance of taking into account both the EW and QCD effects in this relation in the considered case is emphasized. It is noted that the discard of the EW corrections leads to over $10\;{{\rm{GeV}}}$ shift in the difference between the pole and running $t$-quark masses. This magnitude exceeds essentially the modern  uncertainties of the considered relation, following from the treatment of the Tevatron and LHC data where both pole and running $t$-quark masses are defined
in the widespread approach when only the QCD corrections are kept in mind between them.

\end{abstract}

%\pacs{12.38.-t, 12.15.Lk, 11.10.Hi}

\newpage

\section{Introduction}
\pagenumbering{arabic}
\pagestyle{plain}
\setcounter{page}{1}

The top-quark mass is the important theoretical parameter, which is extracted from experimental data of the Tevatron and LHC (see e.g. \cite{Zyla:2020zbs}
 and reviews \cite{Nason:2017cxd, Corcella:2019tgt, Hoang:2020iah}). Among the various definitions of the renormalized top-quark masses, the pole and running masses are applied more often.
The pole mass $M_t$ is determined in the on-shell (OS) renormalization scheme by position of the pole of the renormalized $t$-quark propagator on the mass shell $p^2=M^2_t$ in the Minkowskian time-like region of energies.
The running scale-dependent mass $\overline{m}_t(\mu^2)$ is defined within the modified minimal-subtraction $\msbar$-scheme of dimensional regularization. 

The pole mass is a finite quantity but it is sensitive to small momenta and, as a result, suffers from the large perturbative corrections \cite{Beneke:1994sw}. This effect is associated with manifestation of the infrared (IR) renormalons \cite{Beneke:1994sw, Bigi:1994em, Beneke:1998ui}. They make the perturbative coefficients of the relation between the pole $M_t$ and running mass $\overline{m}_t(\mu^2)$  factorially growing with increasing order of perturbation theory (PT) in QCD \cite{Beneke:1994sw, Bigi:1994em,  
Beneke:1998ui, Beneke:2016cbu, Pineda:2001zq, Hoang:2017suc, Kataev:2018mob, Kataev:2018gle}.  The given feature leads to the asymptotic behavior of the PT series for the considered relation in QCD. Therefore, in the specific physical studies it is important to know when this asymptotic structure will start to reveal itself. The estimates of the higher orders of PT in QCD to the $M_t/\overline{m}_t(\overline{m}^2_t)$-ratio, performed in \cite{Kataev:2018gle} with help of the results of \cite{Ball:1995ni}, indicate that the first traces of its asymptotic renormalon nature will be manifested from the seventh order of PT only. However,
within the Standard Model (SM) it is also necessary to take into account the electroweak (EW) corrections.
In the OS-scheme they depend functionally on the pole masses of the SM particles, such as the top-quark, $W$ and $Z$-bosons, the Higgs boson. This mass dependence is especially important at the scale $\mu^2\sim M^2_t$, where masses $M_W$, $M_Z$, $M_H$ are equal in order of magnitude to $M_t$. Therefore, the study of the possible asymptotic structure of the $M_t/\overline{m}_t(M^2_t)$-ratio in the EW sector is more difficult than in the case of QCD. Further on, we will see that accounting for these effects may overlap the QCD ones.

In the SM the ratio $M_t/\overline{m}_t(\mu^2)$ contains three types of corrections:
\begin{equation}
\label{M_t-m_t-delta}
\frac{M_t}{\overline{m}_t(\mu^2)}=1+\delta^{{\rm{QCD}}}(\mu^2)+\delta^{{{\rm{EW}}}}(\mu^2)+\delta^{{\rm{QCD}}\times {\rm{EW}}}(\mu^2),
\end{equation}
where $\delta^{{\rm{QCD}}}$, $\delta^{{{\rm{EW}}}}$, $\delta^{{\rm{QCD}}\times {\rm{EW}}}$ are the QCD, EW and mixed QCD-EW PT contributions correspondingly. They have the following form:
\begin{gather}
\delta^{{\rm{QCD}}}(\mu^2)=\sum\limits_{n\geq 1}\delta^{{\rm{QCD}}}_n(\mu^2)\bigg(\frac{\alpha_s(\mu^2)}{4\pi}\bigg)^n, ~~~\delta^{{\rm{EW}}}(\mu^2)=\sum\limits_{n\geq 1}\delta^{{\rm{EW}}}_n(\mu^2)\bigg(\frac{\alpha(\mu^2)}{4\pi\sin^2\theta_W(\mu^2)}\bigg)^n, \\
\delta^{{\rm{QCD}}\times {\rm{EW}}}(\mu^2)=\sum\limits_{n\geq 2}\sum\limits_{k=1}^{n-1}\delta^{{\rm{QCD}}\times {\rm{EW}}}_{k, n-k}(\mu^2)\bigg(\frac{\alpha_s(\mu^2)}{4\pi}\bigg)^k\bigg(\frac{\alpha(\mu^2)}{4\pi\sin^2\theta_W(\mu^2)}\bigg)^{n-k}. 
\end{gather}

Here $\alpha_s(\mu^2)=g^2_s(\mu^2)/4\pi$,  $\alpha(\mu^2)=e^2(\mu^2)/4\pi$; $g_s$ and $e$ are the coupling constants of the $SU(3)_c$ and $U(1)_{em}$ gauge groups correspondingly, $\sin\theta_W(\mu^2)$ is the sine of the Weinberg angle. All these parameters are determined here in the $\msbar$-scheme. In this case the scale-dependent coefficients $\delta^{{\rm{EW}}}_n(\mu^2)$, $\delta^{{\rm{QCD}}\times {\rm{EW}}}_{k, n-k}(\mu^2)$ are the functions depending on the running top-quark mass, $W$, $Z$ and Higgs boson masses. Naturally, all these corrections may be rewritten through mass and coupling parameters determined in the OS-scheme.
 
Nowadays the QCD corrections $\delta^{{\rm{QCD}}}_1$, $\delta^{{\rm{QCD}}}_2$, $\delta^{{\rm{QCD}}}_3$ and $\delta^{{\rm{QCD}}}_4$ are known due to direct calculations, performed in \cite{Tarrach:1980up}, \cite{Gray:1990yh, Avdeev:1997sz},  \cite{Melnikov:2000qh, Chetyrkin:1999qi} and \cite{Marquard:2016dcn} respectively. The pure EW coefficients $\delta^{{\rm{EW}}}_1$ \cite{Hempfling:1994ar} and $\delta^{{\rm{EW}}}_2$  \cite{Kniehl:2014yia, Kniehl:2015nwa}  were obtained in the analytical and semi-analytical form correspondingly. The mixed two-loop QCD-EW correction $\delta^{{\rm{QCD}}\times {\rm{EW}}}_{1,1}$ is known from analytical computations carried out in \cite{Jegerlehner:2003py, Jegerlehner:2012kn}.
 
As has already been pointed out in the works \cite{Jegerlehner:2003py, Jegerlehner:2012kn, Kniehl:2015nwa}, the EW effects are not negligible for the relation between the pole and running top-quark masses within the most natural from our point of view {\it{Fleischer-Jegerlehner}} (FJ) {\it{scheme}} \cite{Fleischer:1980ub}. In this scheme the vacuum expectation value $v(\mu^2)$ of the Higgs field is renormalized and it includes in the $\msbar$ running mass accordingly to relation $\overline{m}_t(\mu^2)=y_t(\mu^2)v(\mu^2)/\sqrt{2}$ (here $y_t(\mu^2)$ is the running top-quark Yukawa coupling). Moreover, in the one-loop approximation the EW correction to $M_t/\overline{m}_t(M^2_t)$ is larger than the $\mathcal{O}(\alpha_s)$ QCD contribution, which has the opposite sign: $\vert \delta^{{{\rm{EW}}}}_1(M^2_t)\vert \gtrsim \vert \delta^{{\rm{QCD}}}_1(M^2_t)\vert$. The statement about the importance of taking the EW effects into account for different physical quantities
was previously made in many works (see e.g.  \cite{Bochkarev:1994gu, Martin:2016xsp,  Martin:2019lqd, Passarino:2021uxa} and references therein). For instance, the similar situation, when the EW contribution dominates the first non-zero QCD one, was identified in  
\cite{Bochkarev:1994gu} for the $\rho$-parameter of the SM (the ratio of the amplitudes of neutral and charged weak currents at low energies) or more correctly for the value $\Delta\rho=1-1/\rho$ in a wide domain of the Higgs boson mass values from 60\;{\rm{GeV}} to 1\;{\rm{TeV}}. 
 
Despite the importance of the EW effects, we will show that in fact {\it{they are still either underestimated or not taken into account at all
upon the transition from the pole to running top-quark mass}} in the contemporary phenomenologically-oriented literature. The explicit estimates presented below and references to works on this topic, including the results of different collaborations of the Tevatron and LHC and the results given in the PDG, indicate this. Most often, only the QCD effects are kept in mind when the transition from $M_t$ to $\overline{m}_t(\mu^2)$ is realized   (see e.g. the experimental works \cite{Abazov:2011pta, Sirunyan:2018goh, Aad:2019mkw}).  This work is devoted to the discussion and clarification of this problem.

\section{The role of electroweak effects}
\label{Sec2}

\subsection{One-loop analysis}

At the one-loop level the perturbative relation between the pole and running masses of $t$-quark in the SM reads:
\begin{eqnarray}
\label{M_t-m_t-mu}
\frac{M_t}{\overline{m}_t(\mu^2)}=1+\delta^{{\rm{QCD}}}_1(M^2_t/\mu^2)\frac{\alpha_s(\mu^2)}{4\pi}+\delta^{{\rm{EW}}}_1(M^2_t/\mu^2)\frac{\alpha(\mu^2)}{4\pi\sin^2\theta_W(\mu^2)},
\end{eqnarray}
where $\alpha_s(\mu^2)$, $\alpha(\mu^2)$, $\sin^2\theta_W(\mu^2)$ are defined in the $\msbar$-scheme. Note that we will not distinguish the pole and running masses of particles in expressions for $\delta^{{\rm{QCD}}}_1(M^2_t/\mu^2)$ and $\delta^{{\rm{EW}}}_1(M^2_t/\mu^2)$ because it would be beyond the considered $\mathcal{O}(\alpha_s)$ and $\mathcal{O}(\alpha)$ accuracy. The same is also true for the scales $\mu^2=M^2_t$ or $\mu^2=\overline{m}^2_t(\overline{m}^2_t)$. For instance, this can be clearly illustrated from the solution of a one-loop RG equation defining the running effect of the strong coupling at six active quark flavors ($n_f=6$):  $\alpha_s(M^2_t)\approx\displaystyle\frac{\alpha_s(\overline{m}^2_t)}{1-7\alpha_s(\overline{m}^2_t)/(4\pi)\ln(\overline{m}^2_t/M^2_t)}\approx \frac{\alpha_s(\overline{m}^2_t)}{1-7\cdot 0.1/(4\pi)\ln(162.5^2/172.4^2)}\approx \frac{\alpha_s(\overline{m}^2_t)}{1+0.0066}\approx \alpha_s(\overline{m}^2_t)$. The number 0.0066 in the denominator corresponds to the $\mathcal{O}(\alpha_s)$-correction to $\alpha_s$. This small effect will be  neglected in this subsection devoted to the one-loop analysis.

Throughout this section we will utilize the pole masses of the particles of the SM contributing to the coefficients $\delta^{{\rm{QCD}}}_1$ and $\delta^{{\rm{EW}}}_1$. We will treat them 
as input and fix accordingly to PDG(20), namely $M_t\approx 172.4\;{{\rm{GeV}}}$, $M_H\approx 125.10\;{{\rm{GeV}}}$, $M_W\approx 80.38\;{{\rm{GeV}}}$, $M_Z\approx 91.19\;{{\rm{GeV}}}$ \cite{Zyla:2020zbs}.

The one-loop $\mathcal{O}(\alpha_s)$ QCD correction to the ratio $M_t/\overline{m}_t(\mu^2)$ was calculated a long time ago in \cite{Tarrach:1980up} and has a following form:
\begin{equation}
\label{zQCD}
\delta^{{\rm{QCD}}}_1=C_F\bigg(4-3\ln \frac{M^2_t}{\mu^2}\bigg).
\end{equation}

In the fundamental representation of the Lie algebra of the gauge group $SU(N_c)$ with $N_c$ colors the quadratic Casimir operator is $C_F=(N^2_c-1)/(2N_c)$. For the particular case of the $SU(3)$ color group, realized in nature, $C_F=4/3$.

The one-loop $\mathcal{O}(\alpha)$ EW correction to the ratio $M_t/\overline{m}_t(\mu^2)$ was computed in \cite{Hempfling:1994ar} almost three decades ago. In this quoted work it was presented in a general form which is valid not only for the top-quark, but also for an arbitrary heavy quark. Since the Higgs boson mass was unknown at that time, this term was calculated there for various possible relations between values of particle masses contributing to $\delta^{{\rm{EW}}}_1$. It was expressed through the sine and cosine of the weak mixing angle, the electric charges of quarks and their third components of weak isospin. We apply this representation
to the case of the $t$-quark in the real domain, where $M_t>M_H/2$, $M_t>M_Z/2$, and express $\delta^{{\rm{EW}}}_1$ via the pole masses of top-quark, gauge bosons and Higgs boson\footnote{Its
 analytical form expanded in the limit $M^2_B/M^2_t\ll 1$ (here $B=W, Z, H$) is in agreement with  results of \cite{Eiras:2005yt}.}. This leads us to the following result: 
\begin{eqnarray}
\label{zEW}
&&\delta^{{\rm{EW}}}_1=\delta^{{\rm{EW}}, \;(0)}_1+\delta^{{\rm{EW, \;(L)}}}_1\ln \frac{M^2_t}{\mu^2}, \\
\label{zEW0}
&&\delta^{{\rm{EW}}, \;(0)}_1=-\frac{23}{72}-\frac{M^2_t}{M^2_W}+\frac{25}{36}\frac{M^2_W}{M^2_t}+\frac{1}{2}\frac{M^2_W}{M^2_H}+\frac{1}{2}\frac{M^2_H}{M^2_W}-\frac{5}{9}\frac{M^2_Z}{M^2_t}+\frac{4}{9}\frac{M^2_Z}{M^2_W} \\ \nonumber
&&~~~+\frac{17}{72}\frac{M^4_Z}{M^2_WM^2_t}
+\frac{1}{4}\frac{M^4_Z}{M^2_WM^2_H}-\boxed{N_c\frac{M^4_t}{M^2_WM^2_H}}+\frac{3}{2}\frac{M^2_W}{M^2_H}\ln\frac{M^2_t}{M^2_W}+\frac{1}{16}\frac{M^4_H}{M^2_WM^2_t}\ln \frac{M^2_t}{M^2_H}
\\ \nonumber
&&~~~-\frac{1}{8}\frac{M^2_t}{M^2_W}\ln \frac{M^2_t}{M^2_t-M^2_W}-\frac{1}{4}\frac{M^2_W}{M^2_t}\bigg(\frac{M^2_W}{M^2_t}-\frac{3}{2}\bigg)\ln\frac{M^2_W}{M^2_t-M^2_W}
+\bigg(\frac{3}{4}\frac{M^4_Z}{M^2_WM^2_H} \\ \nonumber
&&~~~+\frac{2}{9}\frac{M^2_WM^2_Z}{M^4_t}-\frac{5}{18}\frac{M^4_Z}{M^4_t}-\frac{3}{16}\frac{M^4_Z}{M^2_WM^2_t}+\frac{17}{144}\frac{M^6_Z}{M^2_WM^4_t}\bigg)\ln \frac{M^2_t}{M^2_Z}+\frac{M^2_Z}{M^2_t}\bigg(\frac{10}{9}+\frac{5}{9}\frac{M^2_Z}{M^2_t}
\\ \nonumber
&&~~~ -\frac{8}{9}\frac{M^2_W}{M^2_Z}-\frac{7}{72}\frac{M^2_Z}{M^2_W}-\frac{4}{9}\frac{M^2_W}{M^2_t}-\frac{17}{72}\frac{M^4_Z}{M^2_WM^2_t}\bigg)\bigg(\frac{4M^2_t}{M^2_Z}-1\bigg)^{1/2}\hspace{-0.2cm}\arccos\bigg(\frac{M_Z}{2M_t}\bigg)
\\ \nonumber
&&~~~+\frac{1}{8}\frac{M^4_H}{M^2_WM^2_t}\bigg(\frac{4M^2_t}{M^2_H}-1\bigg)^{3/2}\hspace{-0.2cm}\arccos\bigg(\frac{M_H}{2M_t}\bigg), \\ 
\label{zEWL}   
&&\delta^{{\rm{EW, \; (L)}}}_1=\frac{1}{3}+\frac{3}{8}\frac{M^2_t}{M^2_W}-\frac{1}{3}\frac{M^2_Z}{M^2_W}-\frac{3}{8}\frac{M^2_H}{M^2_W}-\frac{3}{2}\frac{M^2_W}{M^2_H}-\frac{3}{4}\frac{M^4_Z}{M^2_WM^2_H}+\boxed{N_c\frac{M^4_t}{M^2_WM^2_H}}~.
\end{eqnarray}

The expressions (\ref{zEW0}-\ref{zEWL}) are given
in neglecting really very small contributions, associated with mass effects of leptons ($e,\mu,\tau$) and other lighter quarks ($u,d,s,c,b$)  \cite{Hempfling:1994ar}.

Using the abovementioned values of the pole masses $M_t$, $M_W$, $M_Z$, $M_H$, taken from PDG(20), we arrive to the numerical form of the correction $\delta^{{\rm{EW}}}_1$\footnote{It is consistent with other ones, following from the results of \cite{Jegerlehner:2003py, Jegerlehner:2012kn}.}:
\begin{equation}
\label{EW-numerical}
\delta^{{\rm{EW}}}_1=-24.26+25.80\ln\frac{M^2_t}{\mu^2}.
\end{equation}

\begin{figure}[h!]
\centering
\includegraphics[width=0.3\textwidth]{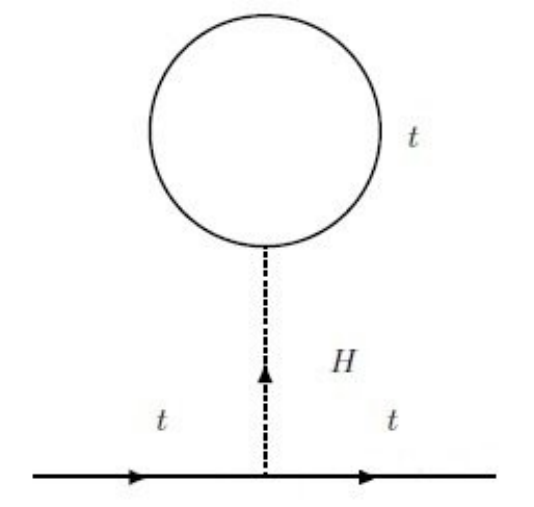}
\captionsetup{justification=centering}
\caption{\label{diagram} One-loop vacuum diagram providing the greatest contribution to the coefficient $\delta^{{\rm{EW}}}_1$.} 
\end{figure}

Pay attention to one interesting fact. The largest contributions in expressions (\ref{zEW0}-\ref{zEWL}) arise due to the tadpole diagrams and the most significant of them are framed. They originate from the diagram presented at the Figure \ref{diagram}
and are equal to the boxed terms $N_cM^4_t/(M^2_WM^2_H)\approx 26.21$, which almost completely form the total values of the coefficients $\delta^{{\rm{EW}}, \;(0)}_1$ and $\delta^{{\rm{EW, \; (L)}}}_1$ (see Eq.(\ref{EW-numerical})). They contain three enhancing multipliers, namely the number of colors $N_c=3$, the fractions $M^2_t/M^2_W\approx 4.6$ and $M^2_t/M^2_H\approx 1.9$.
The incorporation of the tadpole diagrams within the FJ-scheme in a set of self-energy Feynman diagrams of theory with a broken gauge symmetry is necessary to preserve gauge invariance of the relation between $M_t$ and $\overline{m}_t(\mu^2)$ \cite{Hempfling:1994ar,  Bochkarev:1994gu, Jegerlehner:2003py, Jegerlehner:2012kn, Kniehl:2015nwa, Fleischer:1980ub, Bezrukov:2012sa}. Variant of the tadpole-free $\msbar$-scheme was investigated in \cite{Martin:2016xsp, Martin:2019lqd}, where it was noted that its application to the study of perturbative relation between the pole and running top-quark masses leads to the gauge dependence of this ratio.

Let us now move on to the expression (\ref{M_t-m_t-mu}), normalized at the scale $\mu^2=M^2_t$. To get the numerical values for $\alpha_s(M^2_t)$, $\alpha(M^2_t)$, $\sin^2\theta_W(M^2_t)$ we will use the following input $\alpha_s(M^2_Z)\approx 0.1179$, $\alpha^{-1}(M^2_Z)\approx 127.952$, $\sin^2\theta_W(M^2_Z)\approx 0.231$ \cite{Zyla:2020zbs} and then will shift them to the top-quark mass scale.

The corresponding RG equations, defining the running effect of $\alpha_s$ and $\alpha$ (see e.g. \cite{Jegerlehner:2002em, Kniehl:2015nwa, Bezrukov:2012sa} and references therein) in the SM in the one-loop approximation, read:
\begin{eqnarray}
\label{RG-couplings}
\mu^2\frac{\partial a_s}{\partial \mu^2}=-\bigg(11-\frac{2}{3}n_f\bigg) a^2_s, ~~~~~ \mu^2\frac{\partial a}{\partial \mu^2}=\bigg(-7+\frac{4}{3}n_G+\frac{20}{27}n_GN_c\bigg)a^2,
\end{eqnarray}
where $a_s=\alpha_s/4\pi$, $a=\alpha/4\pi$, $n_f$ is the number of quark flavors, $n_G$ is the number of fermion generations, $N_c$ is the number of colors. 
In the SM the dependence of $a=e^2/16\pi^2$ on the scale may be determined from the running of the couplings $g$ of $SU(2)_L$ group,  $g'$ of $U(1)_Y$ group and the relation $1/e^2=1/g^2+1/g'^{\;2}$. The contribution $4n_G/3$ in (\ref{RG-couplings}) corresponds to the QED $\beta$-function in the one-loop approximation; the term $20n_GN_c/27$ is the contribution of the up- and down-type quarks in the $n_G$ generations; the term $-7$ is the effect of  interaction of the photon with $W$-boson.

We fix $n_f=5$, $n_G=3$, $N_c=3$ in the process of  transition from the scale $\mu^2=M_Z^2$ to $\mu^2=M^2_t$. The solutions of Eqs.(\ref{RG-couplings}) have the following form:
\begin{eqnarray}
\label{running couplings}
\alpha_s(\mu^2)\approx \frac{\alpha_s(M^2_Z)}{1-\displaystyle \frac{23}{3}\frac{\alpha_s(M^2_Z)}{4\pi}\ln\frac{M^2_Z}{\mu^2}}, ~~~~ 
\alpha(\mu^2)\approx \frac{\alpha(M^2_Z)}{1+\displaystyle\frac{11}{3}\frac{\alpha(M^2_Z)}{4\pi}\ln\frac{M^2_Z}{\mu^2}}.
\end{eqnarray}

Employing the foregoing input for $\alpha_s(M^2_Z)$ and $\alpha^{-1}(M^2_Z)$ and using Eqs.(\ref{running couplings}), we get that $\alpha_s(M^2_t)\approx 0.108$, $\alpha^{-1}(M^2_t)\approx  127.58$. 

The shift of the $\msbar$-scheme sine of the weak mixing angle from the initial normalization at $\mu^2=M^2_Z$ to $\mu^2=M^2_t$ is implemented by us with the help of the evolution formula, obtained in \cite{Erler:2004in}:
\begin{equation}
\label{runAngle}
\sin^2\theta_W(M^2_t)\approx \frac{\alpha(M^2_t)}{\alpha(M^2_Z)}\sin^2\theta_W(M^2_Z)+\frac{21}{44}\bigg(1-\frac{\alpha(M^2_t)}{\alpha(M^2_Z)}\bigg)+\frac{\alpha(M^2_t)}{4\pi}\bigg(\frac{625}{132}\ln\frac{M^2_t}{M^2_Z}+\frac{18}{11}\ln\frac{\alpha(M^2_t)}{\alpha(M^2_Z)}\bigg).
\end{equation}

Application of this formula gives the approximate value $\sin^2\theta_W(M^2_t)\approx 0.234$ (in comparison with initial input value $\sin^2\theta_W(M^2_Z)\approx 0.231$). 

Utilizing now the expressions (\ref{zQCD}), (\ref{EW-numerical}) in conjunction with the values for $\alpha_s(M^2_t)$, $\alpha^{-1}(M^2_t)$, $\sin^2\theta_W(M^2_t)$ defined above, we arrive to the following relation between the pole and running top-quark masses, obtained 
within the FJ-scheme at the one-loop level:
\begin{equation}
\label{one-loop-M_t}
\frac{M_t}{\overline{m}_t(M^2_t)}\approx 1+\stackrel{\mathcal{O}(\alpha_s)}{\strut{0.046}}-\stackrel{\mathcal{O}(\alpha)}{\strut{0.065}} \;= 0.981 < 1 \;(!?)
\end{equation}

As seen the one-loop EW correction dominates the QCD one and its incorporation in the analysis changes the sign of the total one-loop contribution to the ratio (\ref{one-loop-M_t}). This effect was also previously detected in the work \cite{Jegerlehner:2012kn} (see Figure 1 and Table 1 of this cited paper) and in \cite{Kniehl:2015nwa} (see corresponding Table 1) at $M_H=125\;{\rm{GeV}}$. 

The similar situation may be revealed for the decay width of the Higgs boson in two photons from the results of Ref.\cite{Davies:2021zbx}, where the four-loop QCD correction was first computed  and the two-loop EW one was presented (see original works \cite{Degrassi:2005mc, Actis:2008ts}, where this EW term was calculated). As follows from these manuscripts, the two-loop EW contribution is very close in magnitude to the two-loop QCD one (and it is significantly larger than the three- and four-loop QCD corrections) and has the opposite sign to it.

Let us return to the discussion of the $M_t/\overline{m}_t(M^2_t)$-ratio (\ref{one-loop-M_t}). Its numerical form is in agreement with the results of the works \cite{Jegerlehner:2012kn, Kniehl:2015nwa}.
Note that the magnitude of the $\mathcal{O}(\alpha)$ EW contribution in the difference between $M_t$ and $\overline{m}_t(M^2_t)$ (or between $M_t$ and $\overline{m}_t(\overline{m}^2_t)$\footnote{As was clarified above, at the one-loop level $\overline{m}_t(M^2_t)\approx \overline{m}_t(\overline{m}^2_t)$ in Eq.(\ref{M_t-m_t-mu}).}) is about 10\;{\rm{GeV}}. It is clear that this large correction exceeds considerably all existing uncertainties of the parameters (couplings, $\sin^2\theta_W$, masses of particles), contained 
in Eqs.(\ref{M_t-m_t-mu}), (\ref{zEW0}-\ref{zEWL}).

In its turn, it follows from the definite results of various collaborations, for instance, D0 (see e.g. \cite{Abazov:2011pta}), CMS (see e.g. \cite{Sirunyan:2018goh}) and ATLAS (see e.g. \cite{Aad:2019mkw}), that the currently presented relation between $M_t$ and $\overline{m}_t$ appears to be controlled by the QCD corrections {\it{only}} without taking into account the EW effects\footnote{Indeed, the 3-loop QCD effects are kept in mind only in the ratio between the pole and running top-quark masses in the works \cite{Abazov:2011pta} (see formula (3)) and  \cite{Sirunyan:2018goh} (see page no.23), where the RunDec code \cite{Chetyrkin:2000yt}, based on accounting for the QCD corrections only, is utilized. Whereas, the one-loop QCD correction is merely considered in \cite{Aad:2019mkw}  (see conclusion of this quoted paper and footnote no.9).}. This fact causes some wariness and raises definite questions about values of the scale-dependent top-quark masses obtained from experimental data of the LHC and Tevatron, about their presentation in the current issue of the PDG group and about accounting for the EW effects in the $M_t/\overline{m}_t(\mu^2)$-ratio. 

Consider now the scale evolution of the QCD and EW corrections in Eq.(\ref{M_t-m_t-mu}) above the  threshold top-quark production, namely in the region $M_t\leq \mu\leq 2M_t$. In order not to go beyond the one-loop approximation, we should bear in mind the scale dependence of the coefficients $\delta^{{\rm{QCD}}}_1$ (\ref{zQCD}) and $\delta^{{\rm{EW}}}_1$ (\ref{EW-numerical}) only. The running effects of other parameters, included in Eq.(\ref{M_t-m_t-mu}), are neglected. For a more plausible estimate, further we will fix $\alpha_s$, $\alpha$ and $\sin^2\theta_W$ at the top-quark mass scale, i.e. we will assume that $\alpha_s(\mu^2)\approx\alpha_s(M^2_t)$, $\alpha(\mu^2)\approx \alpha(M^2_t)$, $\sin^2\theta_W(\mu^2)\approx\sin^2\theta_W(M^2_t)$ at $M_t\leq \mu\leq 2M_t$. 

\begin{figure}[h!]
\vspace{-0.5cm}
\centering
\includegraphics[width=1\textwidth]{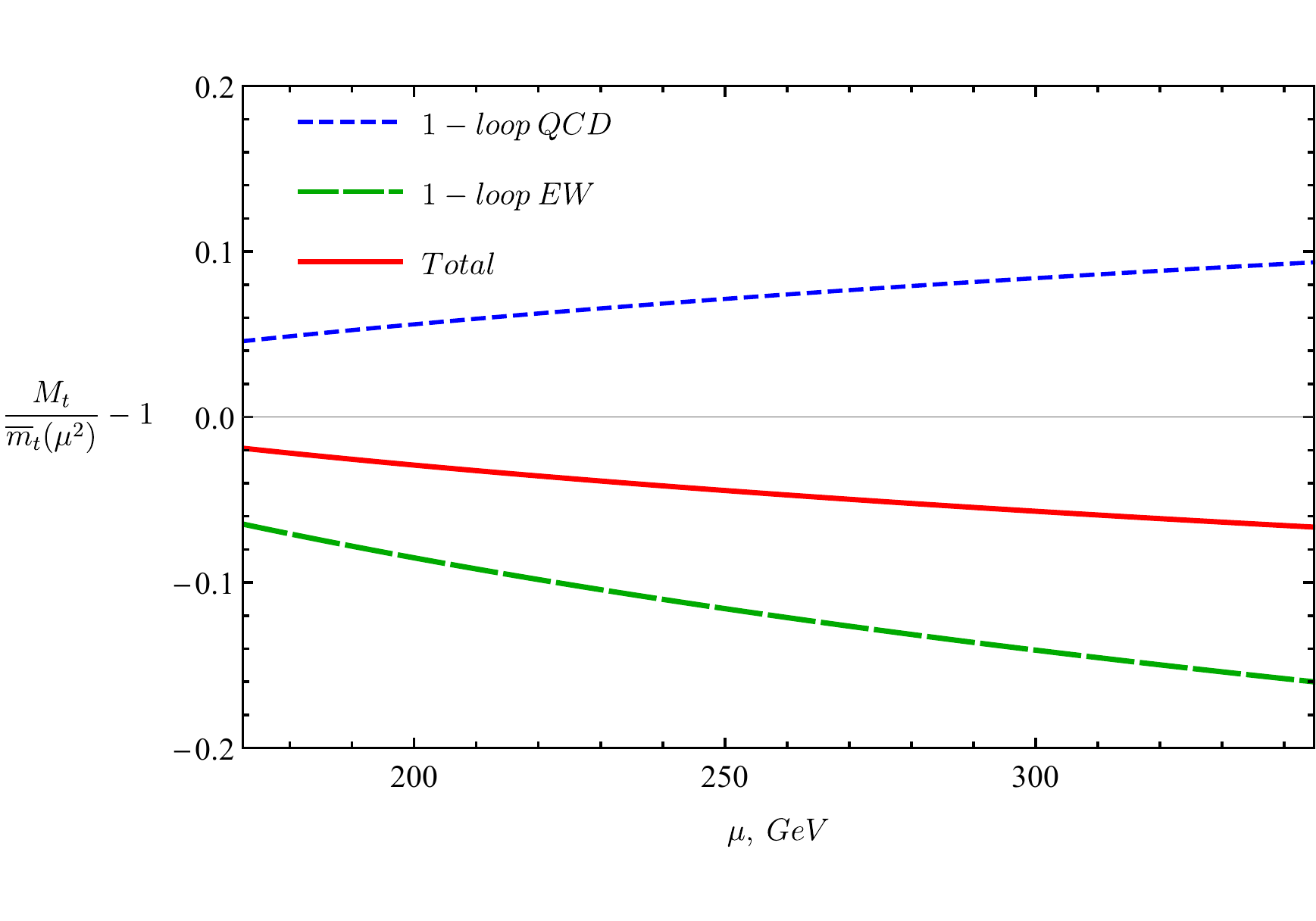}
\vspace{-1cm}
\captionsetup{justification=centering}
\caption{\label{gr2} One-loop QCD and EW corrections to $M_t/{\overline{m}_t(\mu^2)}-1$ at $M_t\leq \mu\leq 2M_t$. } 
\end{figure}

Neglecting now the threshold effects and using the values for $\alpha_s(M^2_t), \alpha^{-1}(M^2_t)$, $\sin^2\theta_W(M^2_t)$, presented above, and expressions (\ref{zQCD}), (\ref{EW-numerical}), we plot the Figure \ref{gr2}, demonstrating the scale evolution of the ratio $M_t/\overline{m}^2_t(\mu^2)$ at the one-loop level in the SM.

It is seen from this figure that the one-loop EW correction dominates the QCD one over the whole domain $M_t\leq \mu\leq 2M_t$. Moreover, it is opposite in sign to the QCD correction. Therefore, the total one-loop correction turns out to be negative. 

One should also mention that the one-loop EW contribution 
$\delta^{{\rm{EW}}}_1$ (\ref{EW-numerical}) may {\it{formally}} be nullified at the scale $\mu^2_0=M^2_t\exp(-0.94)\approx (108\;{\rm{GeV}})^2$.
At this scale we can consider the QCD contribution only, which will be equal to $\delta^{{\rm{QCD}}}_1\approx 1.6$. It is interesting to note that if one leaves in (\ref{zEW0}-\ref{zEWL}) only the leading tadpole contribution coming from the diagram, presented at  Fig.\ref{diagram}, then the scale $\mu^2_0$ will move to the very close value $\tilde{\mu}^2_0=M^2_t\exp(-1)$. 
However, both $\mu_0$ and $\tilde{\mu}_0$ lie below the top-quark mass scale, where the short-distance running mass $\overline{m}_t(\mu^2)$
is not usually used.

Among the low-scale generalizations of the $\msbar$-scheme running mass the concept of the MSR-scheme heavy quark mass $m^{{\rm{MSR}}}_q(R^2)$ \cite{Hoang:2008yj, Hoang:2017suc} is one of the most popular nowadays. The MSR-mass may be evolved to the renormalization scale $R$ below mass of the considered heavy quark (in our case of the top-quark). In practice, this possibility is realized in QCD due to the linear dependency on $R$ in the difference between the OS pole and the MSR-mass: $M_q-m^{{\rm{MSR}}}_q(R^2)=R\sum\limits_{n\geq 1}\delta^{{\rm{MSR}}}_n a^n_s(R^2)$.

We propose to consider the following OS-MSR top-quark mass relation at the one-loop level, thereby extending its QCD definition \cite{Hoang:2017suc} to the case of the SM and requiring the same linear dependency on $R$:
\begin{equation}
\label{R}
M_t-m^{{\rm{MSR}}}_t(R^2)=R\bigg(\frac{16}{3}\frac{\alpha_s(R^2)}{4\pi}-24.26\frac{\alpha(R^2)}{4\pi\sin^2\theta_W(R^2)}\bigg),
\end{equation}
where the coefficients $16/3$ and $-24.26$ are the $\log$-independent terms in the expressions (\ref{zQCD}) and (\ref{EW-numerical}) correspondingly. Note that the tadpole contributions do not disappear in Eq.(\ref{R}), they are still included in the $\log$-independent EW coefficient $-24.26$.

Using now the expression (\ref{R}), the formulas (\ref{running couplings}) at the scale $\mu^2=R^2=\mu^2_0$, defined above, and the relation (\ref{runAngle}) with $R^2=\mu^2_0$ instead of $M^2_t$, we obtain:
\begin{equation}
\label{MSR}
\frac{M_t}{m^{{\rm{MSR}}}_t(\mu^2_0)}\approx 1+\stackrel{\mathcal{O}(\alpha_s)}{\strut{0.03}}-\stackrel{\mathcal{O}(\alpha)}{\strut{0.04}}.
\end{equation}

Thus, even at $\mu_0\approx 108\;{\rm{GeV}}$
the one-loop EW correction, determined in the OS-MSR mass ratio (\ref{MSR}) within the aforesaid approach, exceeds the one-loop QCD one. Note in passing that both of these corrections are slightly smaller than those incorporated in the $M_t/\overline{m}_t(M^2_t)$-ratio (\ref{one-loop-M_t}).

\subsection{Higher-order estimates}
 
Remind that by now not only the one-loop corrections in Eq.(\ref{M_t-m_t-delta}) have been calculated, but the two- \cite{Gray:1990yh, Avdeev:1997sz}, three- \cite{Melnikov:2000qh, Chetyrkin:1999qi}, four-loop \cite{Marquard:2016dcn} QCD and two-loop mixed $\mathcal{O}(\alpha\alpha_s)$ QCD-EW \cite{Jegerlehner:2003py, Jegerlehner:2012kn} and two-loop $\mathcal{O}(\alpha^2)$ EW \cite{Kniehl:2014yia, Kniehl:2015nwa} corrections as well. Taking them into account, one can get the following \textit{approximate expression} for $M_t/\overline{m}_t(M^2_t)$ in the FJ-scheme at the $t$-quark pole mass scale\footnote{At the scale $\mu^2=\overline{m}^2_t(\overline{m}^2_t)$ the QCD part of the relation between $M_t$ and $\overline{m}_t$ reads: $M_t/\overline{m}_t(\overline{m}^2_t)\approx 1+0.046+0.010+0.003+0.001$, where we fix the running $\msbar$-scheme mass accordingly to the PDG(20), namely $\overline{m}_t(\overline{m}^2_t)=162.5\;{\rm{GeV}}$. The corresponding value $\alpha_s(\overline{m}^2_t)=0.1084$ of the strong coupling is defined from its four-loop inverse logarithmic representation with incorporation the matching transformation conditions (for details see e.g. \cite{Kataev:2018gle}). The
 QCD part of Eq.(\ref{two-loop-M_t}) at $\mu^2=M^2_t$
 differs {\textit{a bit}} from the one, presented here at $\mu^2=\overline{m}^2_t(\overline{m}^2_t)$.}:
\begin{equation}
\label{two-loop-M_t}
\frac{M_t}{\overline{m}_t(M^2_t)}\approx 1+\stackrel{\mathcal{O}(\alpha_s)}{\strut{0.046}}+\stackrel{\mathcal{O}(\alpha^2_s)}{\strut{0.013}}+\stackrel{\mathcal{O}(\alpha^3_s)}{\strut{0.004}}+\stackrel{\mathcal{O}(\alpha^4_s)}{\strut{0.002}}
-\stackrel{\mathcal{O}(\alpha)}{\strut{0.065}}+\stackrel{\mathcal{O}(\alpha\alpha_s)}{\strut{0.002}}+\stackrel{\mathcal{O}(\alpha^2)}{\strut{0.005}}\;=1.007 ~(!?)
\end{equation}

The approximate values of the $\mathcal{O}(\alpha\alpha_s)$ and $\mathcal{O}(\alpha^2)$ terms follow directly from the results of Refs.\cite{Jegerlehner:2003py, Kniehl:2015nwa}. Since our input for the pole mass of the top-quark distinguishes slightly from those, used in \cite{Jegerlehner:2003py, Kniehl:2015nwa}, the discussed two-loop contributions are also distinct a bit from their counterparts obtained there.

The expression (\ref{two-loop-M_t}) demonstrates that
 the sum of the one- and two-loop EW and mixed QCD-EW corrections is almost totally compensated by the first four QCD ones. Note that unlike the $\mathcal{O}(\alpha)$ term, the $\mathcal{O}(\alpha\alpha_s)$ and $\mathcal{O}(\alpha^2)$ contributions are positive
 and  are comparable to the $\mathcal{O}(\alpha^3_s)$ correction. 
Thus, the incorporation of the EW effects within the FJ-scheme strongly affects a value of the total correction to the relation between the pole and running top-quark masses in higher orders of PT as well.

One should also mention that the simplified case, when the quantities $M_H, M_t$ are supposed to be much larger than $M_W, M_Z$, 
 was considered in \cite{Kniehl:2014yia, Kniehl:2015nwa}. In this approximation the value 
 of the two-loop EW correction turns out to be 2 times smaller than the genuine one and is compatible with magnitude of the $\mathcal{O}(\alpha^4_s)$ contribution. In addition, an explicit independent calculation \cite{Eiras:2005yt} enables to conclude that not only the one-loop EW term, presented in Eq.(\ref{zEW0}-\ref{zEWL}), but also the two-loop QCD-EW one, evaluated in \cite{Jegerlehner:2003py, Jegerlehner:2012kn}, is in full agreement with its expanded form in the limit $M^2_B/M^2_t\ll 1$ (here $B=W, Z, H$).
This is an extra confirmation of the reliability of the results of the complicated two-loop calculations, performed in \cite{Jegerlehner:2003py, Jegerlehner:2012kn}.

Let us compare the value of the $M_t/\overline{m}_t(M^2_t)$-ratio, given in Eq.(\ref{two-loop-M_t}), with $M_t/\overline{m}_t(\overline{m}^2_t)$ one, following from those treatment of experimental data of the Tevatron \cite{Abazov:2011pta} and LHC \cite{Sirunyan:2018goh, Aad:2019mkw}, where both the pole and $\msbar$-scheme running $t$-quark masses are determined (as was explained in footnote no.5, the different normalizations $\mu^2=M^2_t$ and $\mu^2=\overline{m}^2_t(\overline{m}^2_t)$ will not {\textit{essentially}} change anything). For this purpose, we will briefly describe now how the values of these masses are fixed and how accurate they are.

As is known the pole mass of the $t$-quark is most commonly identified with its Monte Carlo (MC) counterpart (see e.g. descriptions of Refs.\cite{Abazov:2011pta, Aad:2019mkw, Sirunyan:2018goh} and references therein). This MC mass $M^{{\rm{MC}}}_t$ is extracted from the experimental data on the top-quark production in $p\bar{p}$-collisions at the Tevatron (CDF and D0 Collaborations) (see e.g. \cite{TevatronElectroweakWorkingGroup:2016lid, Abazov:2017ktz}) and in $pp$-collisions at the LHC (CMS and ATLAS Collaborations) (see e.g. \cite{Aad:2019mkw, Sirunyan:2018goh}). It is used for comparison of the measured cross sections with the predicted ones by the MC event generators, utilizing parton-shower simulations to model hadronization effects (see reviews \cite{Nason:2017cxd, Corcella:2019tgt, Hoang:2020iah}). The MC mass is determined in such way to be most compatible with experimental data on the invariant mass of the decay products. A conceptual question arises how to link $M^{{\rm{MC}}}_t$-parameter with its top-quark pole mass defined in the SM. Nowadays this issue is extremely relevant since currently the uncertainty of $M^{{\rm{MC}}}_t$ has become smaller than the inaccuracy on its theoretical interpretation, which in order of magnitude is about $0.5\div 1\;{{\rm{GeV}}}$ \cite{Nason:2017cxd, Corcella:2019tgt, Hoang:2020iah, Bachu:2020nqn}\footnote{Note also another interesting issue on the possible link between the currently accepted MC and pole masses of the top-quark with their 
counterpart, which may be determined from the $W^+b$, $W^-\bar{b}$ invariant masses of the $pp\rightarrow W^+W^-b\bar{b}$ process
 and the top-quark width decay (see \cite{Baskakov:2018huw} and references therein).}. 

For instance, the analysis, performed by the ATLAS Collaboration in \cite{Aad:2019mkw} from the juxtaposition of the MC top-quark mass with its pole analog in the  $t\bar{t}+1$-jet process, leads to the following relevant LHC value of $M_t$ with the detailed estimates of the various types of uncertainties: $M_t=171.1 \pm  0.4(stat)$ $\pm 0.9(syst)^{+0.7}_{-0.3} (theor) ~{\rm GeV}$ (compare with the average PDG(20) value $M_t=172.4\pm 0.7\;{\rm{GeV}}$ we have used throughout this work). Here the theoretical uncertainties include the aforesaid inaccuracies in the MC mass interpretation.

Along with the pole mass, it is possible to determine the value of the running top-quark mass \cite{Abazov:2011pta, Sirunyan:2018goh, Aad:2019mkw} by redefinition of $M_t$ by its $\msbar$-scheme analog. 
For example, the present LHC value of the running top-quark mass, obtained by the ATLAS Collaboration \cite{Aad:2019mkw}, reads $\overline{m}_t(\overline{m}^2_t)=162.9\pm 0.5(stat)\pm 1.0(syst)^{+2.1}_{-1.2}(theor)\;{{\rm{GeV}}}$  (in comparison with the average PDG(20) value $\overline{m}_t(\overline{m}^2_t)=162.5^{+2.1}_{-1.5}\;{{\rm{GeV}}}$).

The study of behavior of the cross section $\sigma(t\bar{t})$ in $pp$-collisions allows to fix the value of $\overline{m}_t(\overline{m}^2_t)$ directly \cite{Langenfeld:2009wd, Alekhin:2012py, Fuster:2017rev, Catani:2020tko}. In these works the mentioned cross sections with the SM corrections, initially parameterized through the pole masses of particles, are expressed through the $\msbar$-scheme masses . 

Now, we are ready to find a value of the ratio $M_t/\overline{m}_t(\overline{m}^2_t)$, following from the processing results for the pole and running $t$-quark masses, defined say by the ATLAS Collaboration in \cite{Aad:2019mkw}. Using the foregoing relevant values for these masses, one can obtain that $M_t/\overline{m}_t(\overline{m}^2_t)=1.050^{+0.017}_{-0.012}$ (compare with the one stemming from the PDG(20) values, namely with $M_t/\overline{m}_t(\overline{m}^2_t)=1.061^{+0.014}_{-0.011}$). One can see that within the error bars both these values contradict the one, specified in Eq.(\ref{two-loop-M_t}). This problem is also observed for the $M_t/\overline{m}_t(\overline{m}^2_t)$-ratio, following from the similar analysis of both D0 and CMS  Collaborations (see e.g. \cite{Abazov:2011pta} and \cite{Sirunyan:2018goh}). The reason for this discrepancy is related to the large magnitude of the EW corrections, incorporated in Eq.(\ref{two-loop-M_t}) within the FJ-scheme, but lost in the process of determining the running mass of the top-quark considered above. Let us turn to the discussion of this issue in more detail and clarify where exactly these EW effects are missed.

\section{Discussions}

It follows from Eq.(\ref{two-loop-M_t}) that the one-loop EW contribution in the FJ-scheme is on average 4-6 times greater than the mean-square uncertainties, presented above for the 
$M_t/\overline{m}_t(\overline{m}^2_t)$-ratio, obtained from the definite results of treatment of the LHC data (and also of the Tevatron and PDG(20) ones). Nevertheless, as was already noted, this EW effect appears to be ignored by the members of different collaborations when the transition from the pole to the $\msbar$-scheme mass is carried out in these concrete studies \cite{Zyla:2020zbs, Abazov:2011pta, Sirunyan:2018goh, Aad:2019mkw}.

In the various phenomenologically-oriented works on this topic this effect is not considered when the cross section $\sigma(t\bar{t})$, initially containing the SM corrections and parameterized through the pole masses of particles, is rewritten through the $\msbar$-scheme masses. In process of this transformation {\it{only}} the QCD effects in the relation between $M_t$ and $\overline{m}_t$ are taken into account\footnote{See Eqs.(42-44) of Ref.\cite{Langenfeld:2009wd}; the discussions below Eq.(5) in \cite{Alekhin:2012py}, where the 
 updated analysis of \cite{Langenfeld:2009wd} was performed; Eqs.(9-11) of Ref.\cite{Fuster:2017rev}; Eqs.(8-12) of Ref.\cite{Catani:2020tko}.}. In our opinion, the analysis, undertaken in the Section \ref{Sec2}, points out the investigations, performed in these works, should be supplemented by the inclusion of the EW corrections to the considered top-quark mass relation
as well. As follows from Eq.(\ref{two-loop-M_t}), they will result in a substantial shift in the values of the running top-quark masses obtaining there.

Finally, one should emphasize that, generally speaking, the conception of the $\msbar$-scheme running quark mass may be defined by several ways in the EW sector. Let us briefly discuss the main approaches used in the literature for its definition. In this work we have used the $\msbar$-scheme top-quark mass, defined in the framework of the Fleischer-Jegerlehner tadpole scheme \cite{Fleischer:1980ub} accordingly to the relation $\overline{m}_t(\mu^2)=y_t(\mu^2)v(\mu^2)/\sqrt{2}$ that arises in the spontaneously broken phase.  As was already mentioned above, this scheme enables to preserve the gauge-invariance between the pole and running top-quark mass. Herewith, it is necessary to take into account the tadpole contributions providing this property (see \cite{Hempfling:1994ar, Jegerlehner:2003py, Jegerlehner:2012kn, Kniehl:2014yia, Kniehl:2015nwa}).

Another possible approach, where the relation  $\overline{m}_t(\mu^2)=y_t(\mu^2)v_{eff}(\mu^2)/\sqrt{2}$ is used, leads to the gauge-dependent running mass \cite{Martin:2016xsp, Martin:2019lqd}. Here the effective vacuum expectation value $v_{eff}(\mu^2)$ is determined from the minimization of the effective potential of the Higgs field. In this case the renormalization scheme is a tadpole-free \cite{Martin:2016xsp, Martin:2019lqd}.

One more possible definition is to consider the relation $\overline{m}_t(\mu^2)=y_t(\mu^2)v_0/\sqrt{2}$ with the fixed value $v_0=(\sqrt{2}G_F)^{-1/2}\approx 246\;{\rm{GeV}}$, where the Fermi constant $G_F$ is the coupling for the effective low-energy theory with the four-fermion interaction. Its value is determined usually from the experimental data on the $\mu\rightarrow\nu_{\mu}\bar{\nu}_ee$ decay  \cite{Zyla:2020zbs}. In such gauge-invariant approach the $t$-quark mass runs as the Yukawa coupling $y_t(\mu^2)$ multiplied by the $\mu$-independent factor $v_0/\sqrt{2}$ and it is free of tadpoles \cite{Hempfling:1994ar, Jegerlehner:2012kn, Kniehl:2014yia, Kniehl:2015nwa}. This leads to an order of magnitude smaller EW corrections than for those obtained in the FJ tadpole scheme \cite{Kniehl:2014yia}. However, the absence of the running of the vacuum expectation value is postulated ``by hands'' and the fixed value $v_0$ is not changed during the transition from a low-energy region to a high-energy one, where the top-quark mass is determined.

Thus, nowadays, there is no clear definition of the  running top-quark mass in the EW sector. In this work we have used the first its definition, which, however, leads to the sizeable EW corrections to the relation between $M_t$ and $\overline{m}_t(M^2_t)$, not taken into account in analysis of both Tevatron and LHC data.

\section{Conclusion}

The incorporation of the known two-loop electroweak 
corrections in the analysis of the perturbative relation between the pole and $\msbar$-scheme running masses of $t$-quark 
significantly impacts on the value of the total SM correction between these parameters within the Fleischer-Jegerlehner tadpole scheme, which from our point of view is the most natural among others in the perturbative studies of the SM. It turns out that
the total four-loop QCD and two-loop EW corrections are almost completely canceled out, leading to the relation $M_t\approx \overline{m}_t(M^2_t)$. This fact is in substantial discrepancy with the results of simultaneous determining the pole and running top-quark masses, following from both analysis of the LHC and Tevatron data, where only the QCD effects are kept in mind in the relation between $M_t$ and $\overline{m}_t$. If one takes into account all modern uncertainties, included in definition of these mass parameters, this conclusion will not change. Moreover, the EW effects for the considered ratio are ignored in the phenomenologically-oriented works when the cross section $\sigma(t\bar{t})$, initially parameterized through the pole  mass of the $t$-quark, is rewritten in the terms of the $\msbar$-scheme mass, which is determined numerically afterwards. This causes some wariness and raises definite questions about values of the scale-dependent top-quark masses obtained from the LHC and Tevatron data and about their presentation in the current issue of the PDG group. In fact, these results should be supplemented with the EW corrections (in any of the schemes in which the renormalization of the running mass is determined) and the corresponding extracted mass should be labeled depending on the application of a particular scheme.

\section*{Acknowledgments}

We would like to thank F. Jegerlehner and M.Yu. Kalmykov for useful comments and clarifications which were taken into account in the process of writing  this work. We would also like to express our gratitude to A.B. Arbuzov and especially to N.V. Krasnikov for fruitful and inspiring discussions.
The work of VSM was supported by the Russian Science Foundation, agreement no. 21-71-30003 (investigation of the representation of the one-loop EW correction) and by the Ministry of Education and Science of the Russian Federation as part of the program of the Moscow Center for Fundamental and Applied Mathematics, agreement No. 075-15-2019-1621 (numerical analysis).

\begin{flushleft}

\end{flushleft}

\end{document}